\documentclass[rnote]{aa} 
\usepackage{graphicx}
\usepackage{color}
\usepackage{txfonts}
\def\kms{$\mbox{km s}^{-1}$}

\begin{document}

\title{An updated MILES stellar library and stellar population models}

\authorrunning{Falc\'on-Barroso et al.}
\titlerunning {An updated MILES library and stellar population models}

\author{J. Falc\'on-Barroso\inst{1,2}\fnmsep\thanks{Email: jfalcon@iac.es}, P. S\'anchez-Bl\'azquez\inst{3}, A. Vazdekis\inst{1,2}, E. Ricciardelli\inst{1,2},\\
        N. Cardiel$^{4}$, A.\,J. Cenarro$^{5}$, J. Gorgas$^{4}$, R.\,F. Peletier$^{6}$}

\institute{
$^{1}$Instituto de Astrof\'isica de Canarias. V\'ia L\'actea s/n, La Laguna. Tenerife. Spain\\
$^{2}$Departamento de Astrof\'isica, Universidad de La Laguna, E-38205 La Laguna, Tenerife, Spain\\
$^{3}$Departamento de F\'isica Te\'orica, Universidad Autonoma de Madrid, 28049 Madrid, Spain\\
$^{4}$Dept. de Astrof\'{\i}sica, Fac. de Ciencias F\'{\i}sicas, Universidad Complutense de Madrid, E-28040 Madrid, Spain\\
$^{5}$Centro de Estudios de F\'{\i}sica del Cosmos de Arag\'on, Plaza de San Juan 1, Planta 2$^{\rm a}$, E-44001, Teruel, Spain\\
$^{6}$Kapteyn Astronomical Institute, University of Groningen, Postbus 800, 9700 AV, Groningen, Netherlands}

\date{Received XXX, XXXX; accepted XXX, XXXX}
 
\abstract
{}
{We present a number of improvements to the MILES library and stellar population
models. We correct some small errors in the radial velocities of the stars,
measure the spectral resolution of the library and models more accurately, and
give a better absolute flux calibration of the models.}
{We use cross-correlation techniques to correct the radial velocities of the
offset stars and the penalised pixel-fitting method, together with different
sets of stellar templates, to re-assess the spectral resolution of the MILES
stellar library and models. We have also re-calibrated the zero-point flux level
of the models using a new calibration scheme.}
{The end result is an even more homogeneously calibrated stellar library than
the originally released one, with a measured spectral resolution of
$\sim$2.5~\AA, almost constant with wavelength, for both the MILES stellar
library and models. Furthermore, the new absolute flux calibration for the
spectra excellently agrees with predictions based on independent photometric
libraries.}
{This improved version of the MILES library and models (version 9.1) is 
available at the project's website (http://miles.iac.es).}


\maketitle

\section{Introduction}
The MILES library consists of 985 stars spanning a large range in atmospheric
parameters (S\'anchez-Bl\'azquez et al. 2006, hereafter S06). The spectra were
obtained at the 2.5m Isaac Newton Telescope at the Observatorio del Roque de los
Muchachos and cover the range 3525-7500\AA\ with an originally estimated spectral
resolution of 2.3~\AA\ full-width half maximum (FWHM). A homogenized compilation
of the stellar atmospheric parameters (T$_{\rm eff}$, Log(g), [Fe/H]) for the
stars of this library was presented in Cenarro et al. (2007). Four years after
its release, MILES has proved to be an often used, well calibrated stellar
library that has been extensively used for a wide range of different
applications (Davis et al. 2007; Emsellem et al. 2007; Shields et al. 2007;
Martins \& Coelho 2007; Cenarro et al. 2008; Scelsi et al. 2008; Zuckerman et
al. 2008; Cid Fernandes \& Gonz{\'a}lez Delgado 2010). The stellar library is
also the basis for our empirical models for stellar population synthesis
(Vazdekis et al. 2010, hereafter V10). Both the stellar library and models can
be easily accessed and manipulated through our dedicated website:
http://miles.iac.es.

During the last year, we have had some private communications with different
research groups reporting that some stars in the stellar library showed radial
velocities different from zero. We have explored this question and, indeed,
confirmed these offsets. In this research note, we summarize our solution for
this problem and also the re-assessment of the spectral resolution for the
stellar library and models, which had also been questioned. While sorting out
these concerns, we also addressed the absolute (but not the relative) flux
scaling for our model spectra. Updated versions of both datasets are available
from our website.

\begin{figure*}
\begin{center}
\includegraphics[angle=0,width=\linewidth]{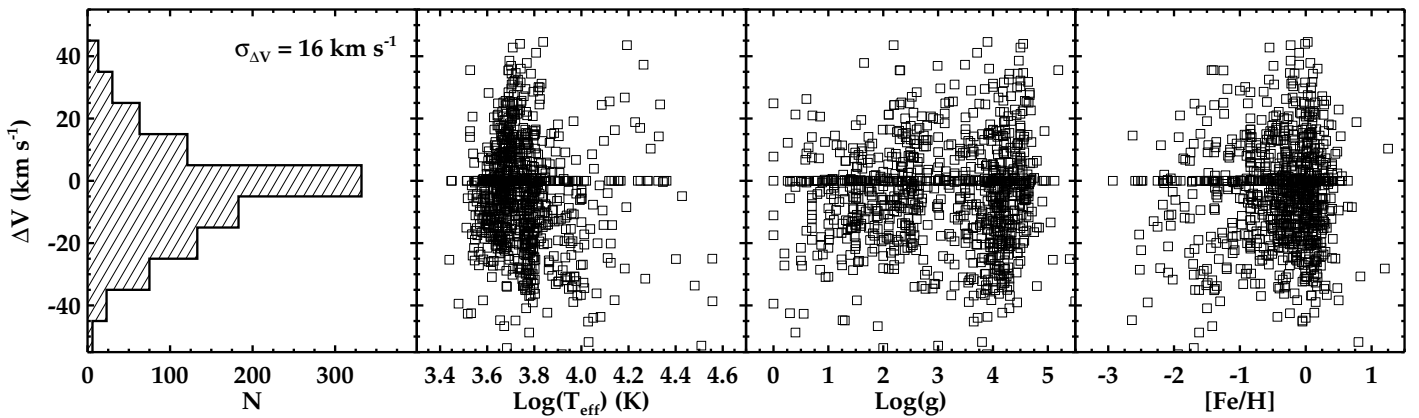}
\caption{Histogram with the mean velocity difference (averaged over the whole
wavelength range) for stars between the originally released MILES stellar
library (v9.0) and the corrected version presented here (v9.1). The dispersion
of the histogram ($\sigma_{\rm \Delta V}$) is also indicated. Offsets are also
presented as a function of atmospheric parameters: effective temperature
(Log(T$_{\rm eff}$)), surface gravity (Log(g)), and metallicity ([Fe/H]).} 
\label{fig:histogram}
\end{center}
\end{figure*}

\section{MILES stellar library}

\subsection{Radial velocity corrections}
\label{sec:velshifts}
Some stars in the originally released version of the stellar library (v9.0) are
not at rest with respect to the Sun. In order to correct from this effect we
cross-correlated each star with a high-resolution solar spectrum from the
BASS2000 database\footnote{http://bass2000.obspm.fr/solar\_spect.php}. This
cross-correlation was performed across wavelength in intervals of 250~\AA\
width. The resulting velocities were fitted, as a function of wavelength, with a
fifth order polynomial, which was used to de-redshift the spectra. The
cross-correlation using the solar spectrum as a template only gives us accurate
velocities for solar-type stars. For the rest, we used as a templates MILES
stars with similar spectral types that were already corrected from the velocity
offsets. We divided the stars into 15 broad groups as a function of their
atmospheric parameters so that only 15 stars were used as templates for the
whole library. This is a compromise between trying to avoid the template
mismatch and the propagation of the errors that result from using one corrected
star as a template to correct another.\looseness-2

Figure~\ref{fig:histogram} shows a histogram with the mean velocity differences
between the MILES stars published in 2006 (v9.0) and the new, corrected ones
(v9.1). The figure indicates that 50\% of the stars in the library were affected
by this problem (at least beyond the expected wavelength calibration
uncertainties, on average 10~\kms). We note and explain below that these offsets
have very little impact on the resulting spectral resolution for our single
stellar population models (hereafter SSPs). The average dispersion for the
sample ($\sigma_{\rm \Delta V}=16$~\kms) is indicated. The figure also shows the
lack of dependence of these offsets on a particular set of stellar parameters.
The list of stars with the corrections will be posted on the MILES
website.\looseness-2

\subsection{Re-assessment of the spectral resolution}
\label{sec:resol}

The data employed to re-assess the resolution of the MILES stellar library 
consists of three sets of spectra (for each version of the library: v9.0 and v9.1):

\begin{enumerate}

 \item \textbf{HD010307}: a star in the MILES stellar library with stellar
parameters (T$_{\rm eff}=5838.0$, Log(g)=4.28, [Fe/H]=0.03) very similar to
those of the Sun, assumed here to have T$_{\rm eff}=5770$, Log(g)=4.4, and
[Fe/H]=0.0.

  \item \textbf{Fake\_Sun}: Best representation of the solar spectrum obtained 
by combining solar-type stars in our library using the interpolation scheme 
described in Vazdekis et al. (2003) for the Sun parameters above.

  \item \textbf{MILES stars}: the 985 spectra from the MILES stellar
library. In order to avoid template mismatch effects derived from limitations of
the different template libraries, we have also selected a subset of 55 stars
with stellar parameters close to those of the Sun (i.e. where most libraries are
assumed to be reliable) for some of the tests.

\end{enumerate}

\noindent The stellar libraries used as templates are

\begin{enumerate}
 \item \textbf{Sun (KPNO)}: Observed solar spectrum obtained from the National
Solar Observatory Atlas (Kurucz et al. 1984). The spectrum covers the wavelength
range between 2970 and 9530~\AA\ with a resolution of $R\sim300000$.

 \item \textbf{Indo-US library}: Empirical library downloaded from
http://www.noao.edu/cflib/. It consists of 1273 stars observed with the 0.9m
Coud\'e Feed telescope at Kitt Peak National Observatory. The spectra typically
cover from 3454 to 9469~\AA, with a nominal spectral resolution (FWHM) of
$\approx$1.2~\AA\ (Valdes et al. 2004). For our tests, only spectra without gaps
in the wavelength regions covered by the MILES dataset were used (193
stars).\looseness-2

 \item \textbf{ELODIE (v3.1) library}: The ELODIE library includes spectra
for 1388 stars obtained with the ELODIE spectrograph at the Observatoire de
Haute-Provence 193cm telescope in the wavelength range 390 to 680 nm. The
library used here is at $R=10000$. Details of the library can be found in 
Prugniel \& Soubiran (2001).

 \item \textbf{S$^4$N library}: The \textit{Spectroscopic Survey of Stars in the
Solar Neighbourhood} (S$^4$N, Allende Prieto et al. 2004) is a library with a
complete spectroscopic catalogue of stars with spectral types brighter than K2V
within 15 parsec from the Sun. The resulting database contains high-resolution
optical spectra for 118 nearby stars (mostly dwarfs) at $R\sim60000$ from 3600 
to 9200~\AA. See http://leda.as.utexas.edu/s4n/ 

 \item \textbf{Coelho theoretical library}: This is the only theoretical library
used in our tests. It contains stellar spectra for different effective
temperatures (3500 $\leq$ Teff $\leq$ 7000K), surface gravities (0.0 $\leq$
log(g) $\leq$ 5.0), metallicities ([Fe/H]$= -0.5, 0.0, 0.2$), and chemical 
compositions ([$\alpha$/Fe] = 0.0 and 0.4). Spectra were retrieved from 
http://www.cruzeirodosul.edu.br/nat/modelos.html, and are described in
Coelho et al. (2005) and Coelho et al. (2007). The models have a nominal 
spectral resolution of 1.0~\AA\ (FWHM) and cover the wavelength range between 
3000~\AA\ and 13000~\AA.

\end{enumerate}

We measured the broadening of the different datasets using the penalized
pixel-fitting (pPXF) method of Cappellari \& Emsellem (2004), with the data and
templates described above. Before running pPXF, both the data and templates were
re-binned to a common velocity scale of 25~\kms. A non-negative linear
combination of those templates, convolved with a Gaussian, was fitted to each
individual test spectrum. The best-fitting parameters were determined by
chi-squared minimization in pixel space. In order to assess the dependence with
wavelength, we divided the original datasets into seven bins. Additionally, a
low-order Legendre polynomial (order 5) was included in the fit to account for
small differences in the continuum shape between the test spectra and the input
templates. 

\begin{figure}
\begin{center}
\begin{minipage}{\linewidth}
\includegraphics[angle=90,width=\linewidth]{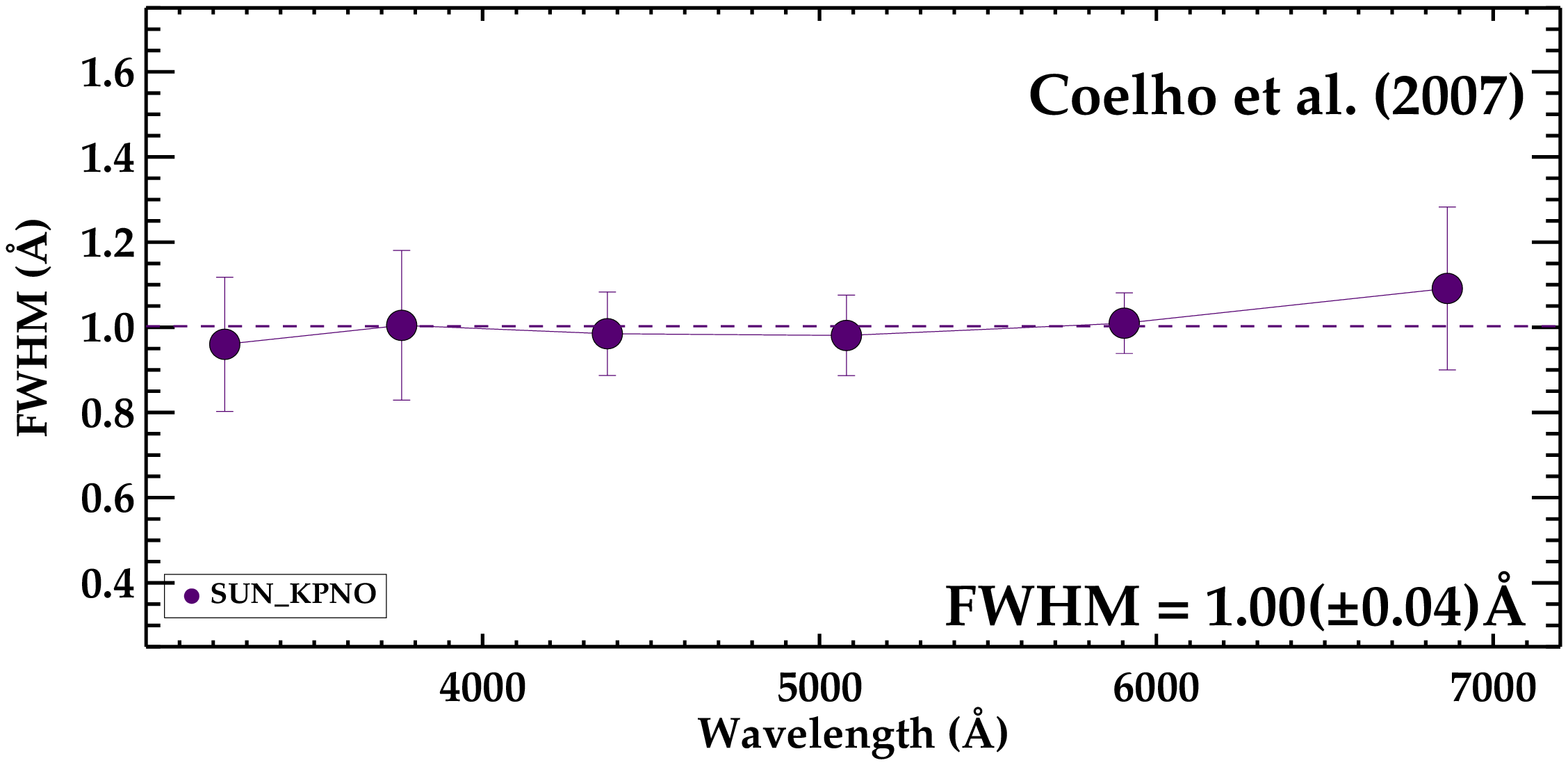}
\vspace{0.15cm}
\end{minipage}
\begin{minipage}{\linewidth}
\includegraphics[angle=90,width=\linewidth]{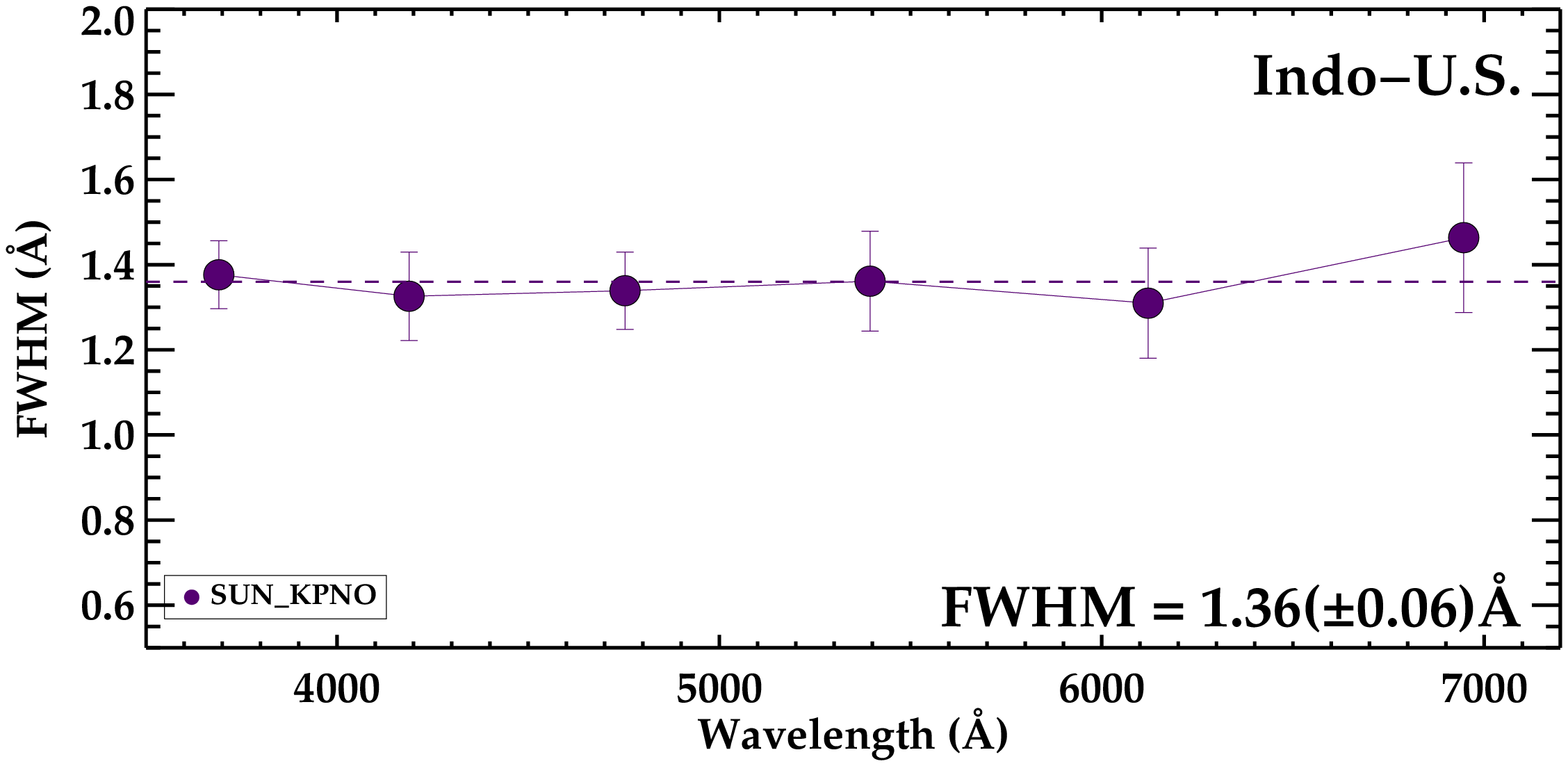}
\end{minipage}
\caption{Measured spectral resolution of the Coelho et al. (2007) and Indo-US 
libraries. A small subset of the libraries (i.e. spectra with parameters close 
to the Sun) were used for this test. We used the solar spectrum from KPNO 
as template.}
\label{fig:coelho}
\end{center}
\end{figure}

Before carrying out the desired tests on the MILES library, we tested whether we
were able to reproduce the quoted spectral resolution of the Indo-US and
Coelho's libraries using the stars with stellar parameters closest to the Sun.
Given that they have FWHM relatively close to the expected values for MILES,
knowing the real value is important to avoid systematic offsets in the measured
resolution. The other libraries have such high spectral resolution that an error
in the exact value has very little impact in our results. The outcome of our
tests are shown in Figure~\ref{fig:coelho}. The plot confirms that the spectral
resolution of Coelho's library is indeed 1.0~\AA, while it shows that for
Indo-US it appears to be slightly higher, FWHM=$1.36(\pm0.06)$~\AA, than the
approximate value estimated by the Indo-US team ($\approx1.2$~\AA). For internal
consistency, we adopt 1.36~\AA\ in our analysis. We note that in S06 we measured
a FWHM of 1.0~\AA\ for the Indo-US library based on much simpler tests. The
difference between the two values is at the centre of the new spectral
resolution quoted in this research note, which is reinforced by the use of other
template stellar libraries.

\begin{figure}
\begin{center}
\begin{minipage}{\linewidth}
\includegraphics[angle=90,width=\linewidth]{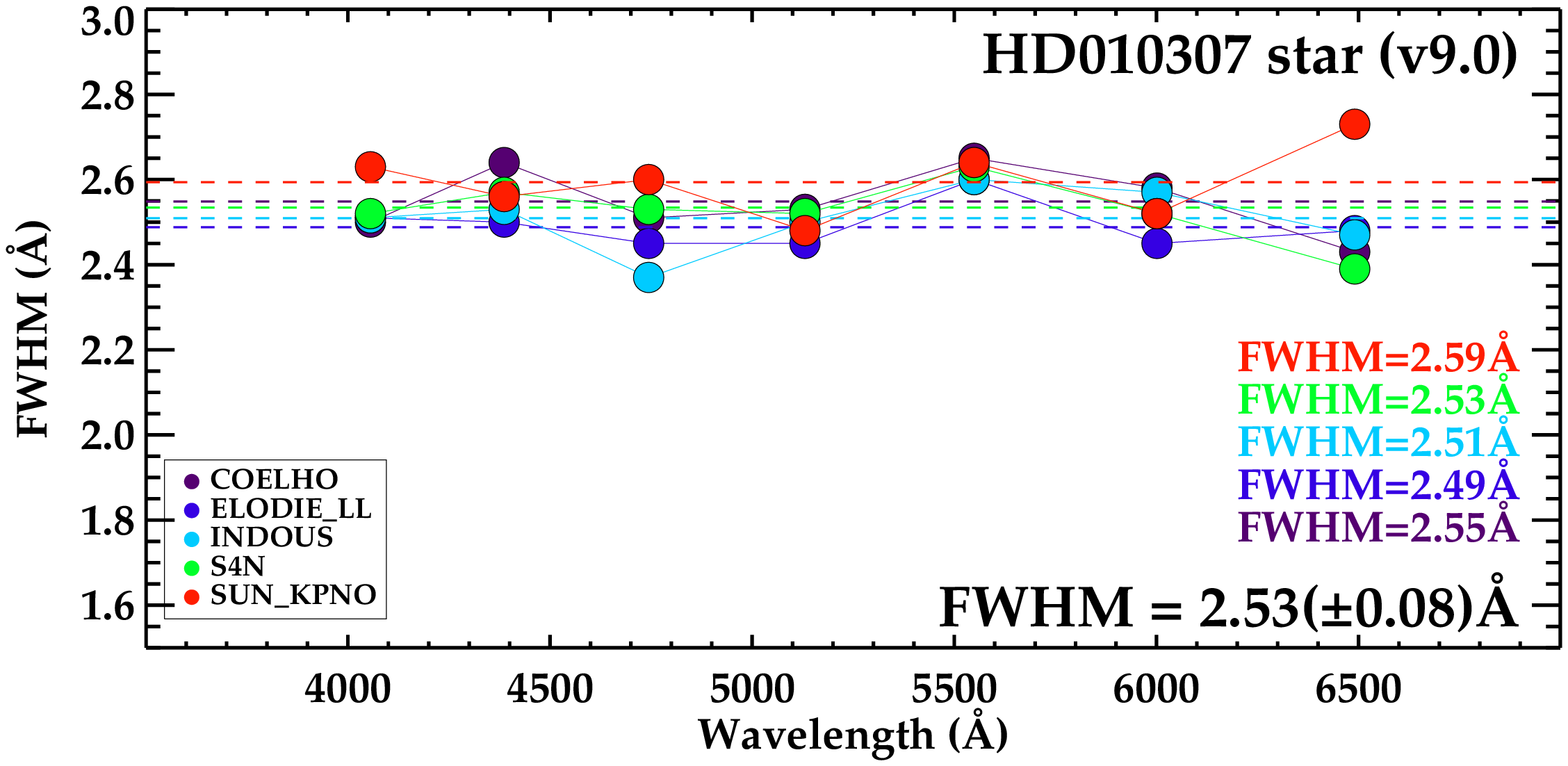}
\vspace{0.15cm}
\end{minipage}
\begin{minipage}{\linewidth}
\includegraphics[angle=90,width=\linewidth]{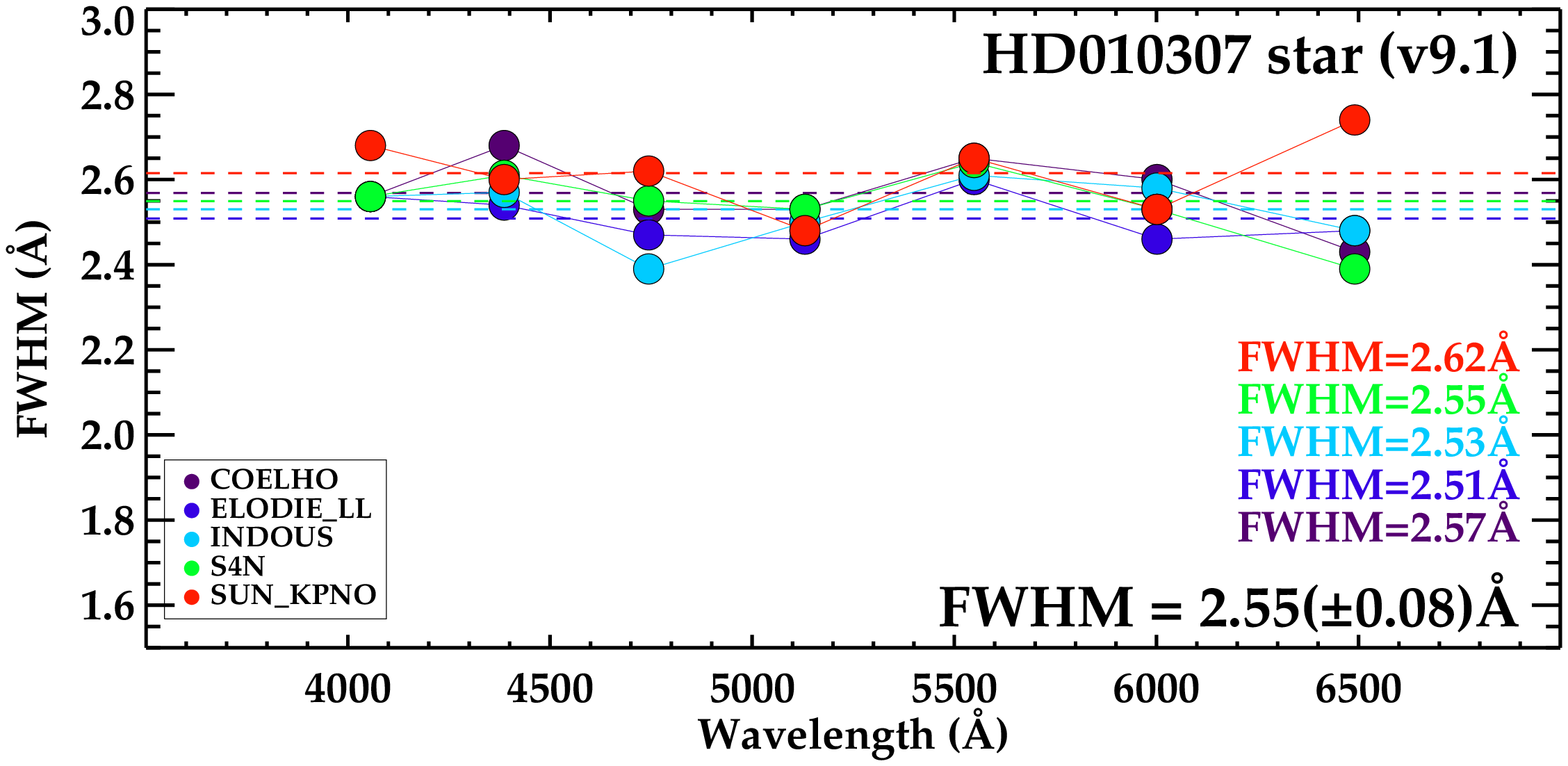}
\end{minipage}
\caption{Measured spectral resolution of the \object{HD010307} spectra for both 
versions. \object{HD010307} is a star in the MILES stellar library with stellar 
parameters: T$_{\rm eff}=5838.0$, Log(g)=4.28, [Fe/H]=0.03).}
\label{fig:m0066v}
\end{center}
\end{figure}

After assessing the true spectral resolution of the Coelho and Indo-US libraries 
we proceeded to determine the resolution of the MILES library using the test 
cases proposed above. These tests were carried out for the original and updated 
versions of the library and models.

\begin{figure}
\begin{center}
\begin{minipage}{\linewidth}
\includegraphics[angle=90,width=\linewidth]{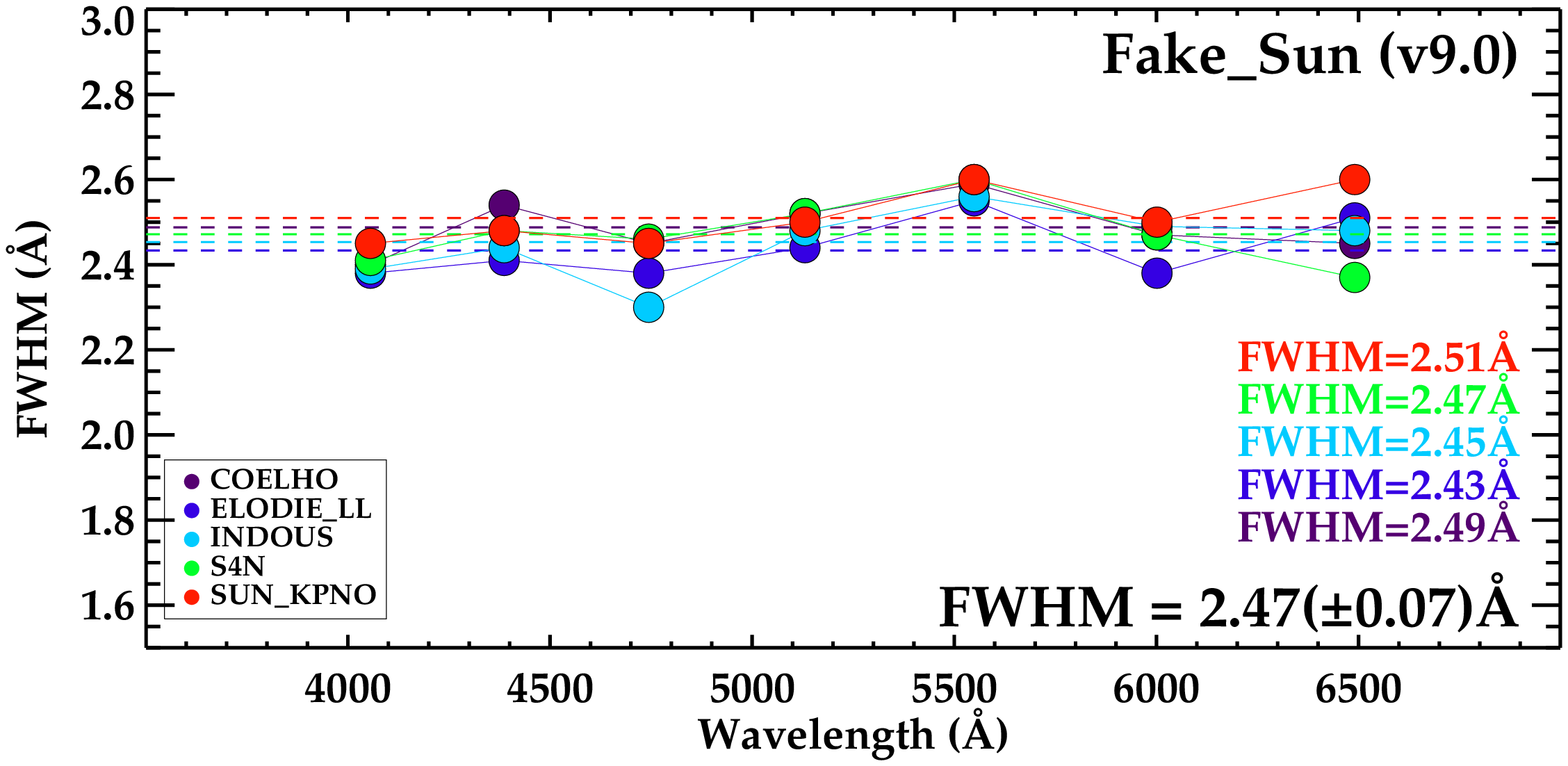}
\vspace{0.15cm}
\end{minipage}
\begin{minipage}{\linewidth}
\includegraphics[angle=90,width=\linewidth]{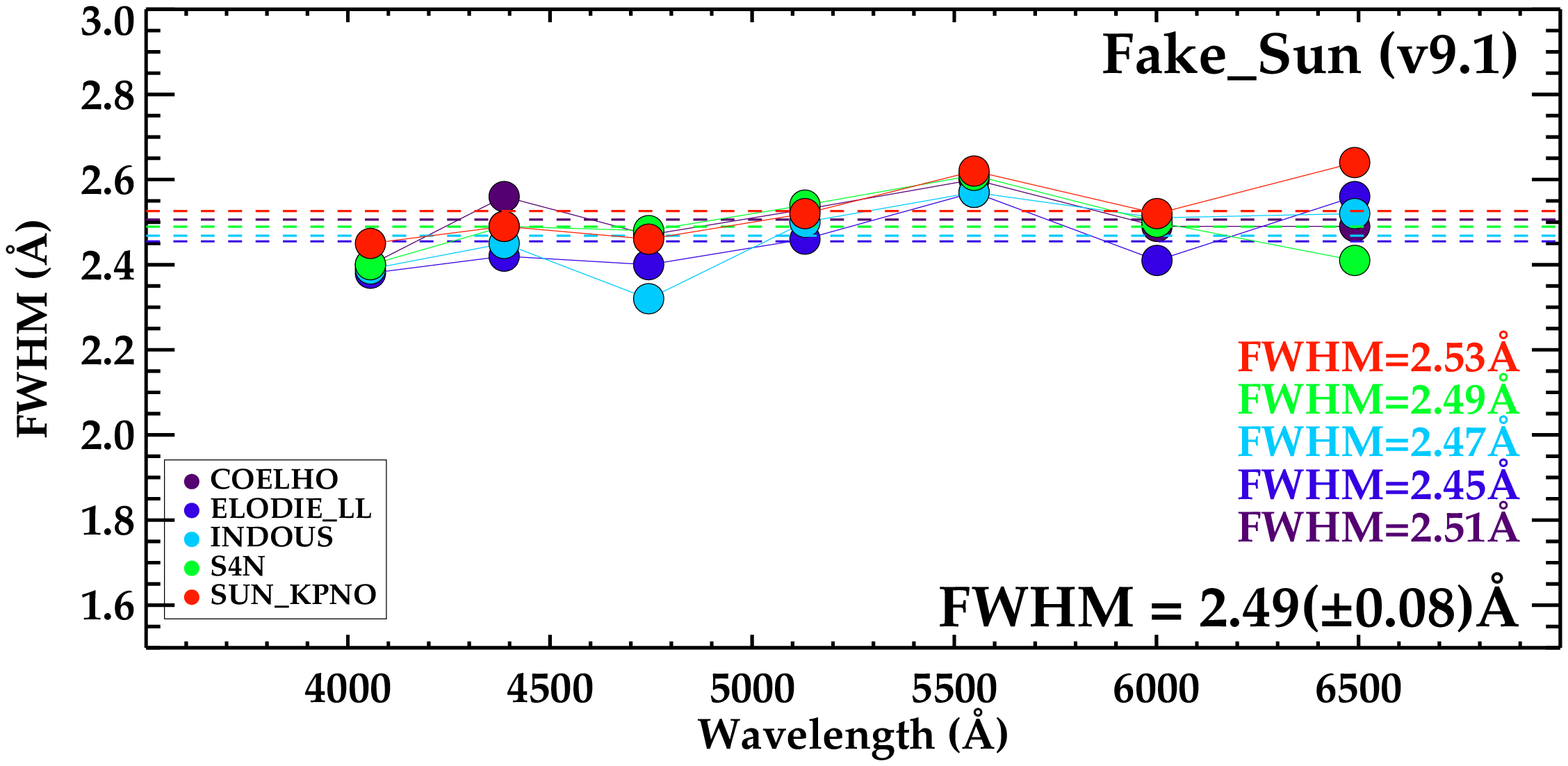}
\end{minipage}
\caption{Measured spectral resolution of the interpolated solar spectrum 
(Fake\_Sun) spectra for both versions. This stellar spectrum is made of the 
following 11 stars in the MILES library: \object{HD076780}, \object{HD186427}, 
\object{HD076151}, \object{HD010307}, \object{HD095128}, \object{HD186408}, 
\object{HD072905}, \object{HD020619}, \object{HD143761}, \object{HD141004}, 
\object{HD200790}.}
\label{fig:fake_sun}
\end{center}
\end{figure}

\begin{figure}
\begin{center}
\begin{minipage}{\linewidth}
\includegraphics[angle=90,width=\linewidth]{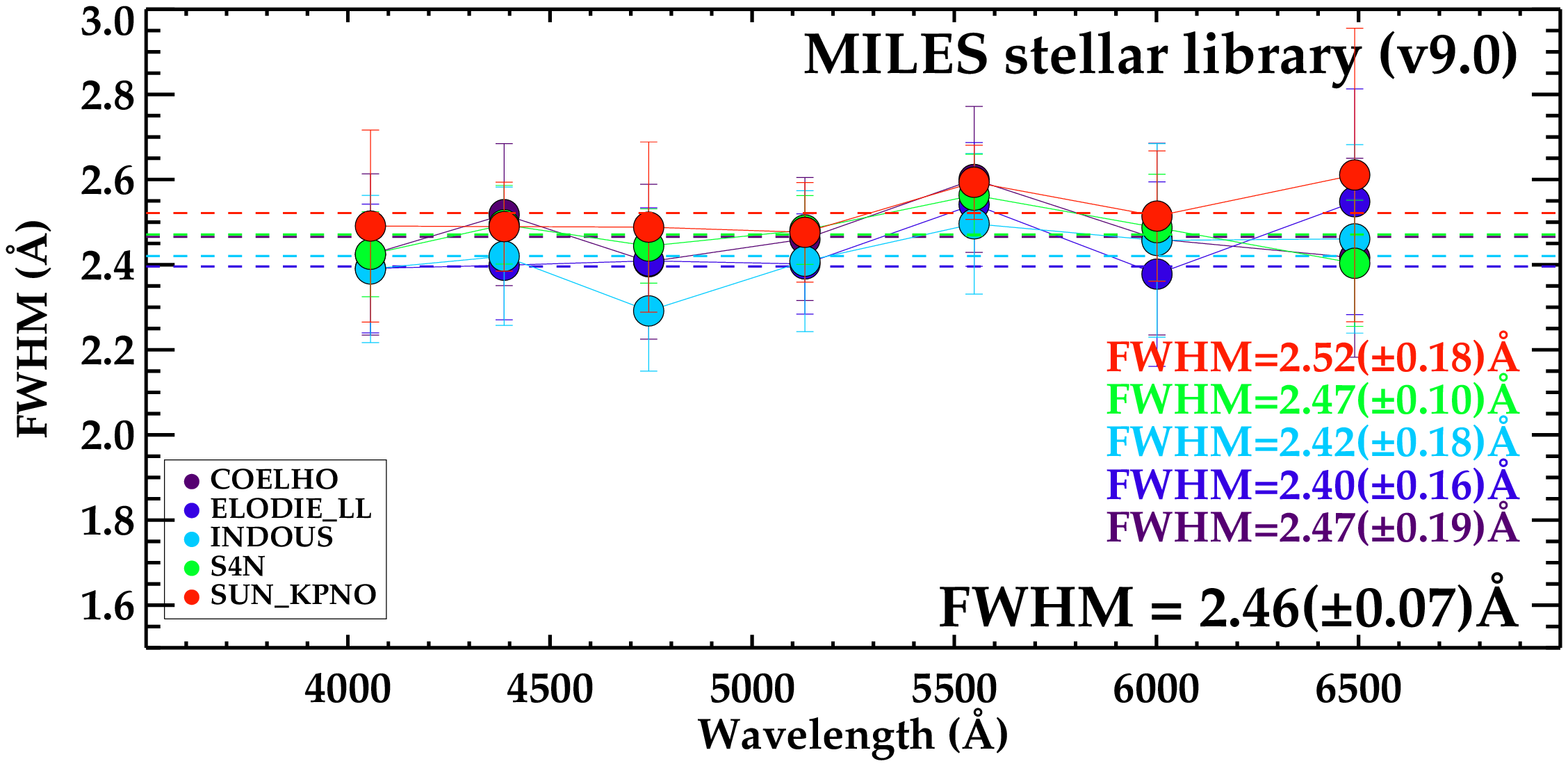}
\vspace{0.15cm}
\end{minipage}
\begin{minipage}{\linewidth}
\includegraphics[angle=90,width=\linewidth]{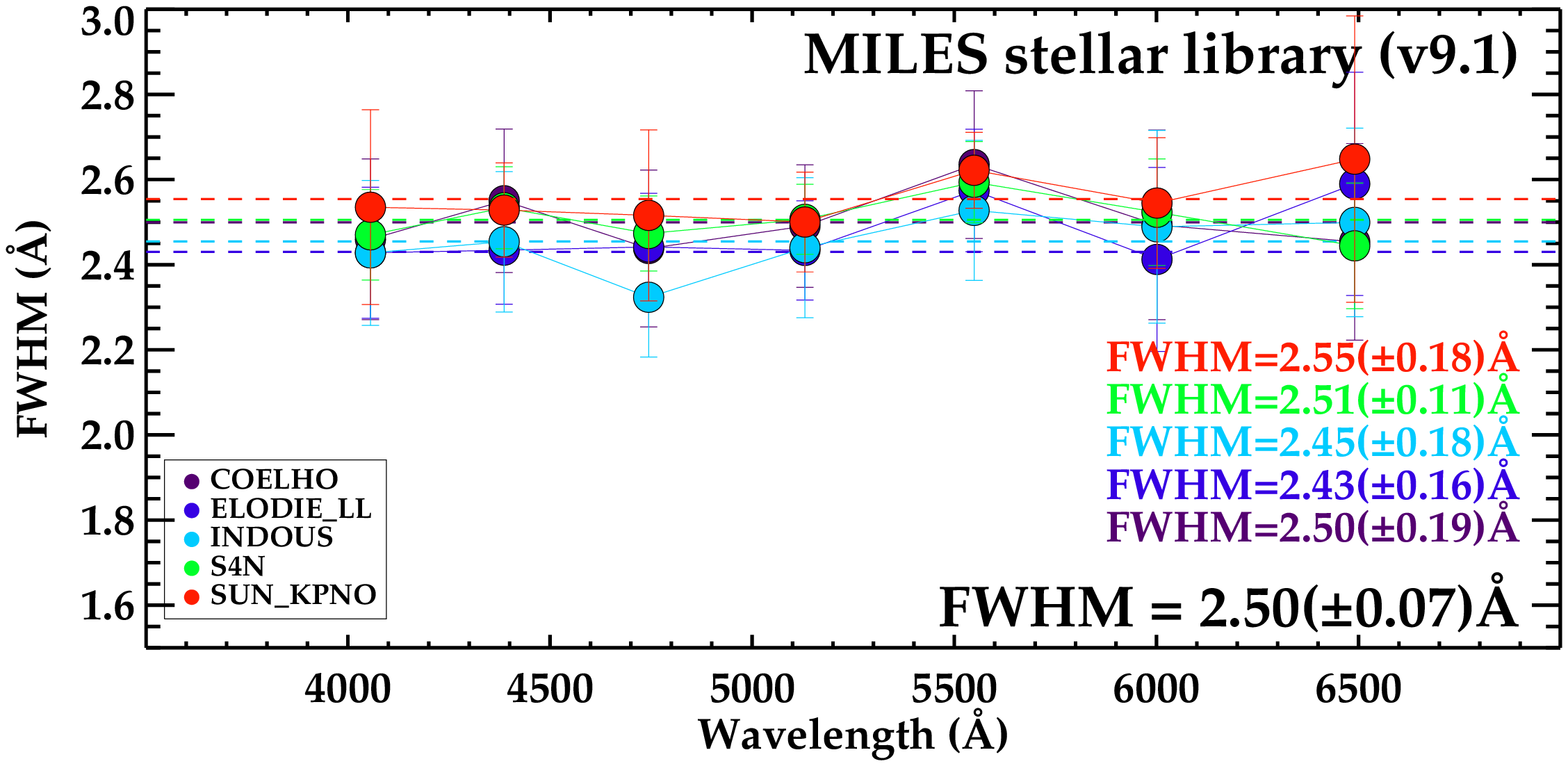}
\end{minipage}
\caption{Measured spectral resolution of the MILES stellar library spectra for 
both versions. The 985 stars in the MILES library were considered when the 
Coelho's, Indo-US, and ELODIE libraries are used as templates. Only stars in 
the library with parameters close to the solar ones were used for the Sun 
(KPNO) and S$^4$N stellar templates. The value in parentheses is the RMS of the 
resolution for each case.}
\label{fig:milesstars}
\end{center}
\end{figure}

\begin{enumerate}
\item \textbf{\textit{MILES single star: \object{HD010307}}}. This is the most
basic test we conducted. In this case we tried to derive the spectral resolution
of the MILES stellar library from a single star. The star was chosen for having
stellar parameters very close to those of the Sun. The results in
Figure~\ref{fig:m0066v} show some scatter as a function of wavelength, where the
regions around H$\beta$ and H$\alpha$ are the noisiest ones. The measured FWHM
is very similar in both versions $\sim$2.55($\pm 0.08$)~\AA. It is worth noting
that results obtained using the Sun as template yields worse results than those
coming from libraries where several templates have been used to derive the best
fit, which highlights the importance of template mismatch in this kind of
tests.\\

\item \textbf{\textit{MILES interpolated solar-type star: Fake\_Sun}}. In this
test we tried to suppress the impact of template mismatch by producing a star as
close as we could to the solar spectrum. We achieved this by means of the
interpolation scheme described in Vazdekis et al. (2003). This spectrum is the
combination of 11 stars in the MILES stellar library. Figure~\ref{fig:fake_sun}
shows the results of our tests. In general, the output values are very similar
to those in the previous test, with the significant difference that the
resulting FWHM has now decreased to $\sim$2.49($\pm 0.08$)~\AA. The comparison
between this and the previous test illustrates the expected level of uncertainty
produced by template mismatch (which is also reflected in our estimated
errors).\\

\item \textbf{\textit{MILES stellar library}}. The previous tests, while very
instructive to assess the order of magnitude of the spectral resolution of the
MILES library, are still incomplete in the sense that they might not reflect the
full range of variations possible in the library. Given the large span in
stellar parameters of the Coelho's, Indo-US, and ELODIE libraries, we used the
complete set of MILES stellar spectra for this exercise. For the Sun (KPNO) and
S$^4$N templates, we selected a small subset of stars of the MILES library with
stellar parameters close to those of the Sun (55 stars) to minimise template
mismatch. The outcome of this test is shown in Figure~\ref{fig:milesstars}. The
results are quite similar to those found for the Fake\_Sun. Individual typical
uncertainties in the measured spectral resolution vary from $0.1$ to $0.2$~\AA.
Also, not surprisingly, the higher FWHM values correspond to those libraries
with the fewer number of templates (e.g. Sun (KPNO) and S$^4$N).

\end{enumerate}

All these tests suggest that the spectral resolution of the new version of the
MILES stellar library is $2.50\pm0.07$~\AA, essentially constant as a function
of wavelength. During the course of this work an independent group carried out
complementary tests and reached similar conclusions (Beifiori et al. 2011).

\subsection{Impact on line-strength indices}
\label{sec:ls_stars}

\begin{figure*}
\begin{center}
\includegraphics[angle=90,width=\linewidth]{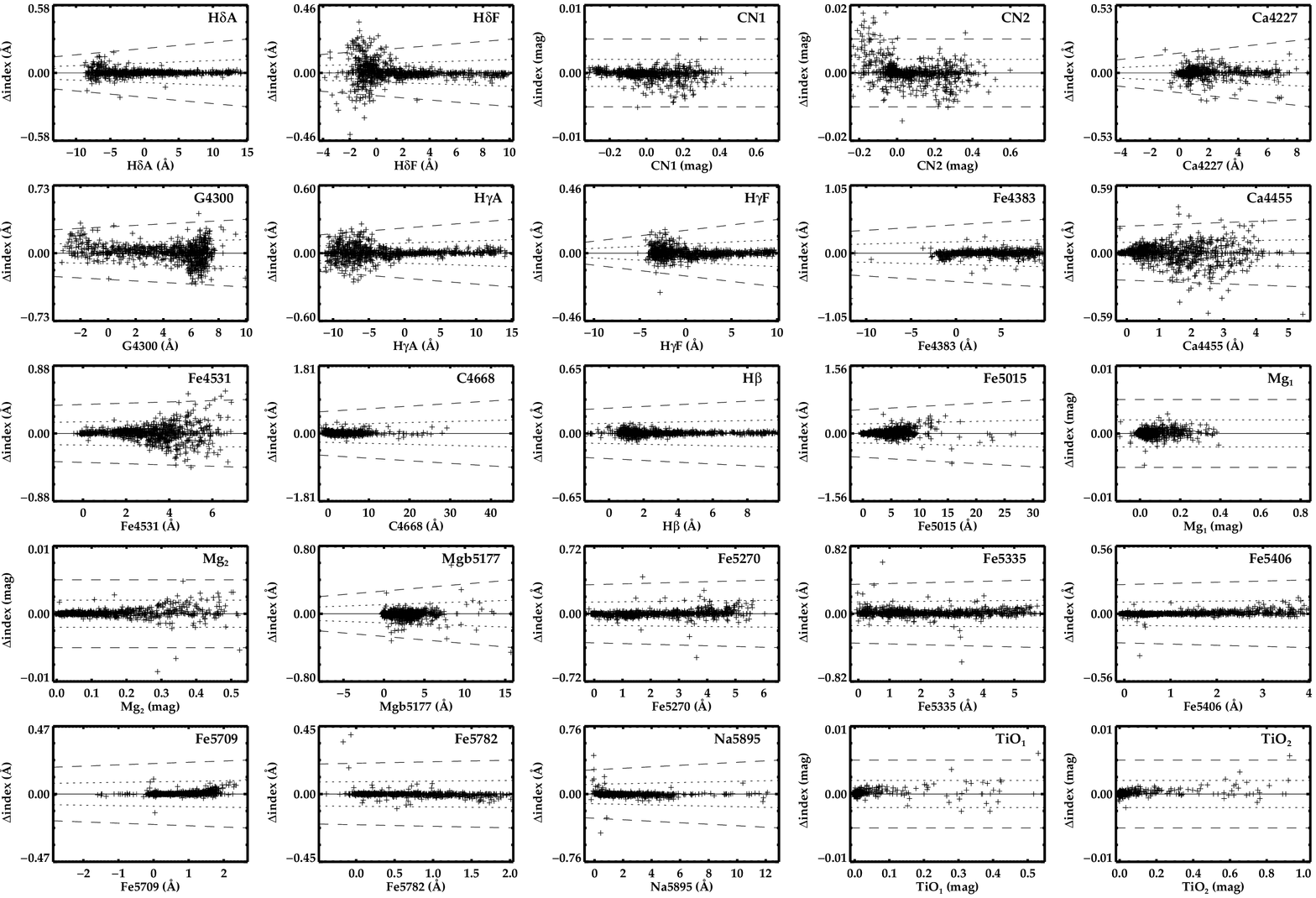}
\caption{Line-strength differences between versions 9.0 and 9.1 of stars in the
MILES stellar library. The solid line marks the zero level. The dashed and
dotted lines indicate the expected uncertainty for spectra with a
signal-to-noise ratio of 30 and 75~\AA$^{-1}$ respectively (see Cardiel et al.
1998). The vertical axis has been adjusted to twice the maximum uncertainty
expected for a spectrum with a signal-to-noise ratio of 30~\AA$^{-1}$ within the
line-strength index range shown.}
\label{fig:miles_ls_stars}
\end{center}
\end{figure*}

Given the shifts in radial velocity applied to the stars of the MILES
library, it is important to assess how these corrections affect the
line-strength indices present in our wavelength range. For this comparison we
concentrated on the Lick indices (e.g. Worthey et al. 1994).
Figure~\ref{fig:miles_ls_stars} shows the difference in the measured indices,
which are in general within the errors of traditional stellar population studies
(e.g. S{\'a}nchez-Bl{\'a}zquez et al. 2006b). For reference, we show in the same
figure the expected uncertainties for index measurements at signal-to-noise
ratios of 30 and 75~\AA$^{-1}$ (Cardiel et al. 1998).

The largest discrepancies appear in line-strength indices at shorter
wavelengths, where the signal-to-noise of the stellar spectra is naturally lower
(S06). The worst cases are H$\delta_{\rm F}$, CN$_2$, G4300, and H$\gamma_{\rm
A}$. This is caused by the overlap of their definition bands (e.g. H$\delta_{\rm
F}$ with CN$_2$, and G4300 with H$\gamma_{\rm A}$). The strongest differences
occur when the Balmer lines are so intense (e.g. at high effective temperatures)
that they start invading the blue and red pseudo-continuum of the CN$_2$ and
G4300 bands respectively. Conversely, the discrepancies at the lowest values of
the H$\delta_{\rm F}$ and H$\gamma_{\rm A}$ indices occur at the highest values
of CN$_2$ and G4300 indices.

\section{MILES stellar population models}

\subsection{Spectral resolution of the MILES models}

In addition to the stellar library, we also carried out the measurement of the
spectral resolution for the MILES single stellar population models. These are
based on the combination of MILES stars in a meaningful way (in the physical
sense) following the prescriptions described in V10. For this test and to
minimize template mismatch, we used Kroupa Universal SSP models with
[M/H]=[$-1.31,-0.71,-0.4,+0.0,+0.22$] and Age=[$0.5,17$] Gyr (160 models). This
time we did not include results from the KPNO and S$^4$N libraries because of
the significant template mismatch observed in those fits (i.e. the solar spectra
did not describe the SSP models properly, and the S$^4$N library is mostly
composed by dwarf stars and thus could not reliably reproduce the SSP spectra,
where giants represent a considerable contribution to the total light in the
MILES spectral range).

\begin{figure*}
\begin{center}
\includegraphics[angle=90,width=0.49\linewidth]{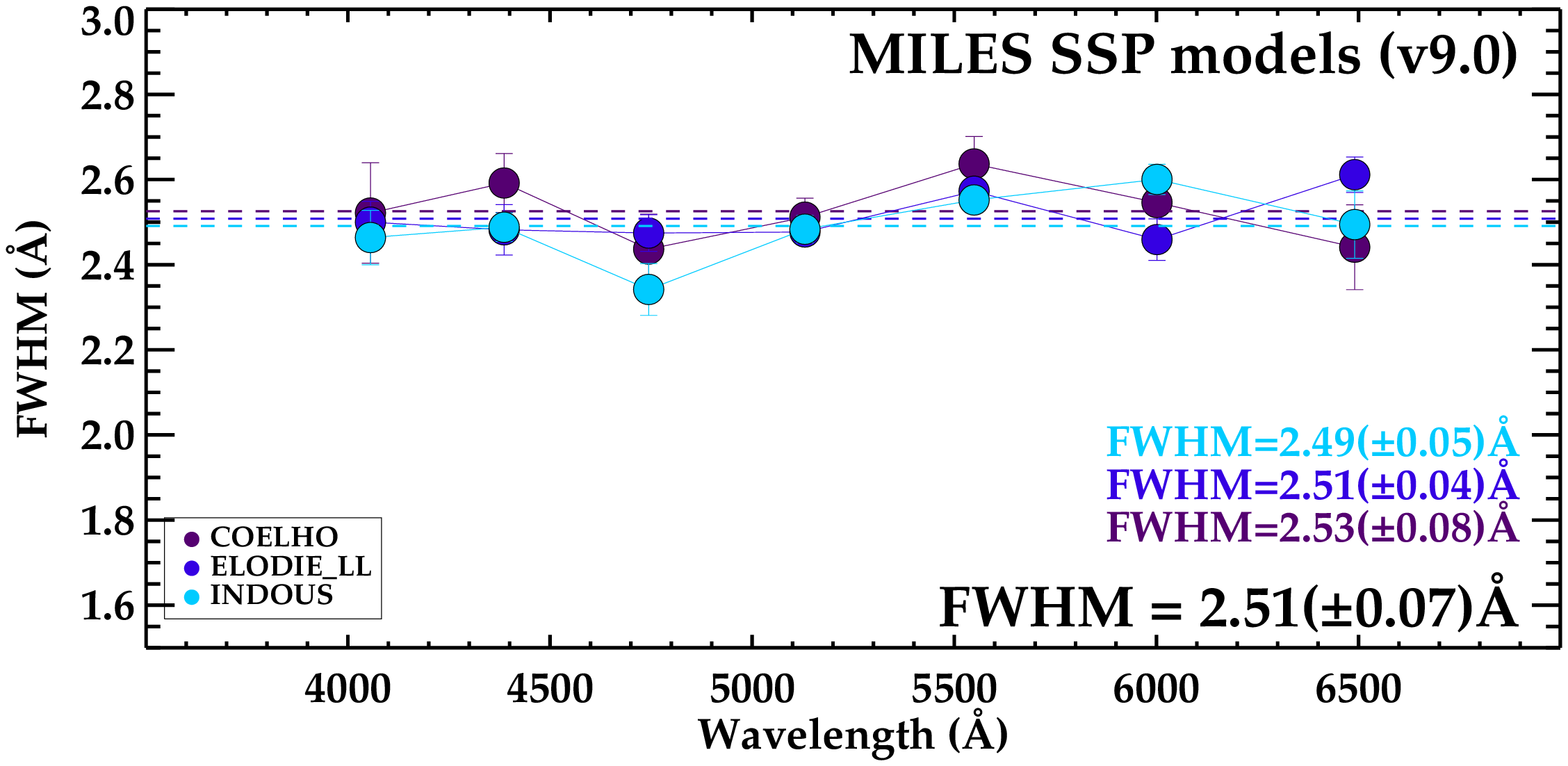}
\hspace{0.15cm}
\includegraphics[angle=90,width=0.49\linewidth]{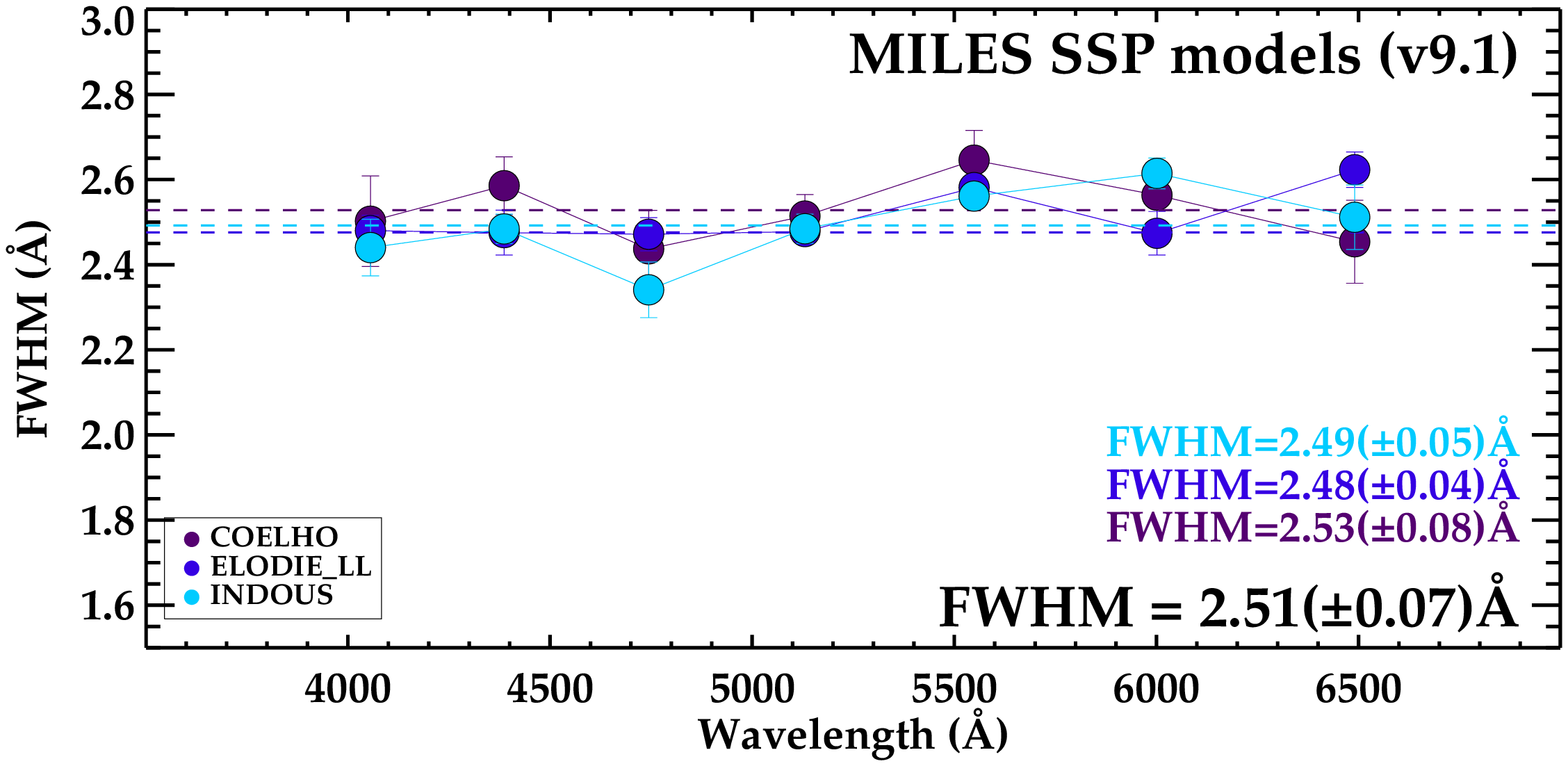}
\caption{Measured spectral resolution of the MILES SSP model spectra for both
versions. To minimize template mismatch, we concentrated on SSP models
with [M/H]$=[-1.31,-0.71,-0.4,+0.0,0.22]$ and Age=[0.5,17] Gyr (160 models). The 
KPNO and S$^4$N spectra were not included in this test because they do not 
contain enough representative stars to accurately reproduce the SSP models.}
\label{fig:milesssp}
\end{center}
\end{figure*}

Given that the models are the mixture of many different types of stars (e.g.
from A to M stars), one could expect the resulting FWHM to be worse than that of
the stellar library. Figure~\ref{fig:milesssp} shows that this is not the case.
The end FWHM is $2.51\pm0.07$~\AA\ and thus very similar to that of the stellar
library. The agreement between the two versions of SSP models seems remarkable,
given the differences in radial velocities of some stars in the stellar library
(see \S\ref{sec:velshifts}). In order to understand this question we calculated
the expected increase or decrease in spectral resolution produced by the
dispersion of the velocity offsets presented in Fig.~\ref{fig:histogram}
($\sigma_{\rm \Delta V}=16$~\kms). It turns out that shifts of that order would
affect our measurements by $\approx$0.01~\AA, which is within the uncertainties
in our determinations, and thus have a very small effect on the final values
reported here. Furthermore, the similarity of the results among libraries
strongly supports our measured FWHM for the Indo-US library. If a FWHM of
1.2~\AA\ is assumed, the values reported in all tests using the Indo-US library
would be lower by $\approx$0.1~\AA.\looseness-2

\subsection{Absolute flux-calibration of the MILES models}

In the updated version of the SSP models (v9.1) we also corrected a flaw
detected in the absolute flux scale of the model spectra that we distributed in
v9.0. The magnitudes measured on the MILES v9.0 SSPs were $\approx0.23$ mag
brighter than those synthesized from our photometric approach, which employs
temperature-gravity-metallicity colour relations derived from extensive
photometric stellar libraries, as described in Vazdekis et al. (1996) and V10.
Models with very steep initial mass functions at the low-mass end (i.e. steeper
than Salpeter), however, show smaller offsets. We note that for most
applications that make use of these spectra there is no significant effect,
except for those inferring mass-to-light ratios directly from the model
spectra.\looseness-2

In order to make the two predictions (i.e. the photometric and spectroscopic)
consistent, we changed our scaling method when integrating the stellar spectra
along the isochrone and followed the same approach as was used to derive the
photometric predictions (Vazdekis et al. 1996). To determine the absolute flux
in the $V$-band corresponding to each star, we adopted the $V$ filter response
of Buser \& Kurucz (1978) and  followed the method described in Fukugita,
Shimasaku \& Ichikawa (1995):

\begin{equation}
F_{V} = 10^{-0.4 (V + Zp_{V} - V_{\alpha Lyr})}
\end{equation}

\noindent where $Zp_{V}$ and $V_{\alpha Lyr}$ are the adopted zero-point and $V$
magnitude for the reference Vega spectrum, respectively. To compute the
zero-point, we used the Hayes (1985) spectrum of Vega with a flux of
$3.44\times10^{-9}$\,erg cm$^{-2}$\,s$^{-1}$\,\AA$^{-1}$ at 5556\,\AA. The $V$
magnitude of Vega is set to 0.03 mag, in concordance with the value adopted by
Alonso et al. (1995), because we mostly use Alonso et al. (1996) and Alonso et
al. (1999) for converting the theoretical parameters of the isochrones to the
observational plane. For the integration of a stellar spectrum as requested by
the isochrone, which comes from our interpolation scheme (see Vazdekis et al.
2003), we normalised the total flux in the $V$-band to unity by convolving with
the same filter response. 

With the new calibration scheme the absolute magnitudes derived directly from
the spectra are fully consistent with those computed from the photometric
libraries well below the observational uncertainties ($\leq0.01$ mag) for all
ages, metallicities, and initial mass functions. This new scaling was also
applied to the SSP models based on the Calcium triplet stellar library (Cenarro
et al. 2001; Vazdekis et al. 2003). It is important to remark that although the
absolute flux scale of the new models has changed with respect to the older
models, this has no impact on the relative flux calibration as a function of
wavelength and thus synthetic colours are not affected. As shown in figure 19 of
V10, the difference between the photometric and the spectroscopic $B-V$ colour
for our reference Vega system is within 0.02 mag for the safe model ranges, with
a dispersion of 0.01 mag for the Kroupa initial mass function.

\subsection{Impact on line-strength indices}

\begin{figure*}
\begin{center}
\includegraphics[angle=90,width=\linewidth]{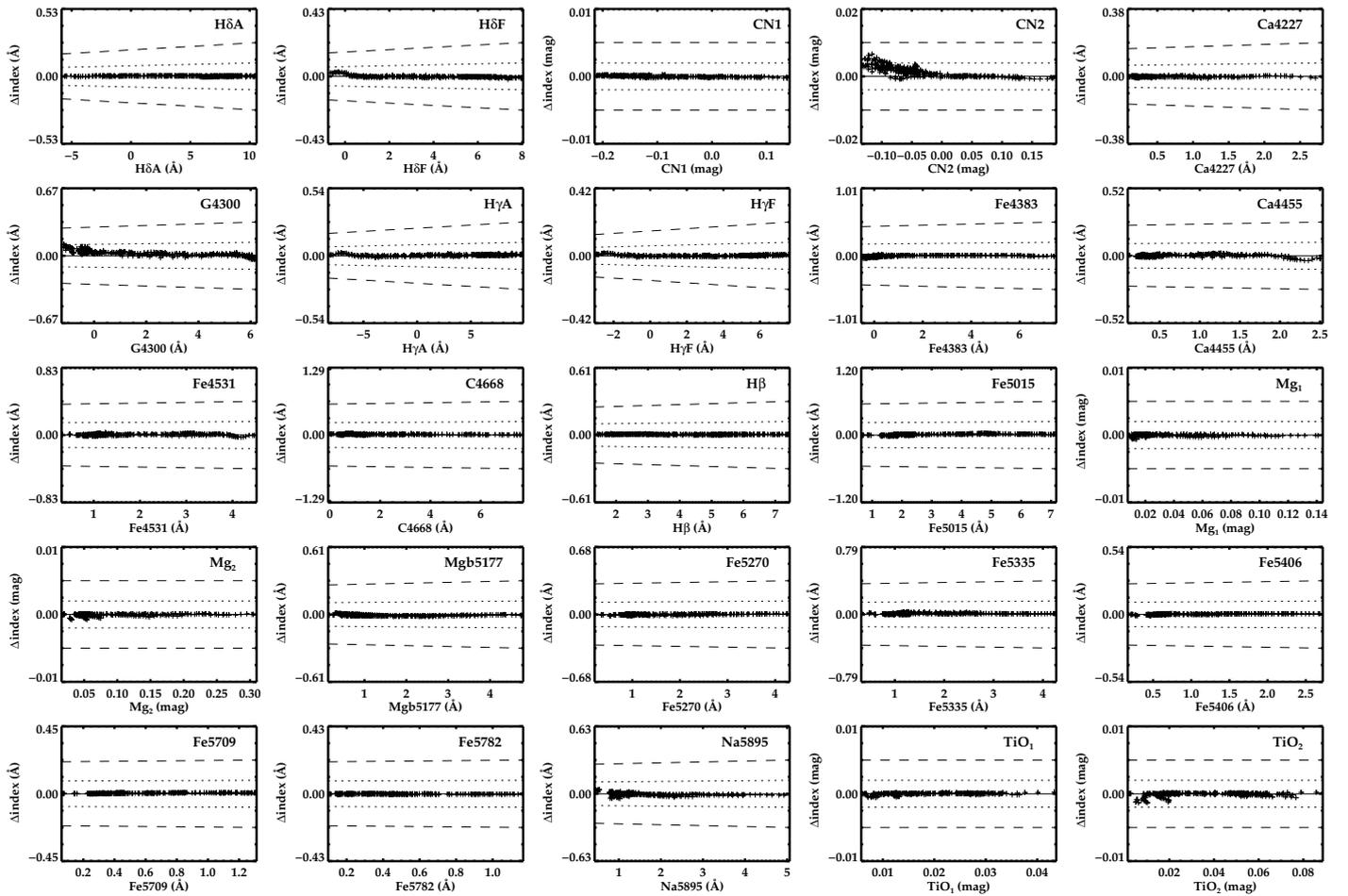}
\caption{Line-strength differences between versions 9.0 and 9.1 of the MILES SSP
models with a Kroupa Universal IMF. Solid line marks the zero level. Dashed and
dotted lines indicate the expected uncertainty for spectra with a
signal-to-noise ratio of 30 and 75~\AA$^{-1}$, respectively (Cardiel et al.
1998). The vertical axis was adjusted to twice the maximum uncertainty expected
for a spectrum with a signal-to-noise ratio of 30~\AA$^{-1}$ within the
line-strength index range shown.}
\label{fig:miles_ls_ssp}
\end{center}
\end{figure*}

Similarly to the stellar library, we have also compared the predicted
line-strength indices for the two versions of the SSP models.
Figure~\ref{fig:miles_ls_ssp} shows this comparison for the Kroupa Universal IMF.
The differences between the two sets of models are naturally much smaller than
those found for the individual stars, given that the models are a combination of
many different types of stars. It is interesting to note the small discrepancies
observed in the CN$_2$ and G4300 indices, which follow the reasons given in
\S~\ref{sec:ls_stars} for the stars. These results show that the impact of the 
radial velocity corrections on the SSP model line-strength indices is generally 
small compared with typical observational uncertainties.

\section{Conclusions}
In this report we summarize our efforts to improve on a number of minor problems
identified in both the MILES stellar library and models. We have
corrected some stars in the stellar library for radial velocity offsets with
respect to the solar rest frame and checked that they have a small impact on the
resulting line-strength indices of the stars and models. We also re-assessed the 
spectral resolution of the MILES library and models, confirming
that it is quite constant as a function of wavelength. We established that
the new values are $2.50\pm0.07$~\AA\ for the stellar library and $2.51\pm0.07$
for the models. We also changed the scheme used to flux calibrate the
spectra in the absolute sense. As a result we find an excellent agreement
($\leq0.01$ mag) between fluxes measured from our spectra and those from
predictions obtained by using photometric (rather than spectroscopic) libraries.
Our tests also show that the spectral resolution of the Indo-US library is
slightly higher (FWHM$=1.36\pm0.06$~\AA) than the approximate value established
by the Indo-US team (FWHM$\approx1.2$~\AA). 

The updated version of the MILES stellar library and models in this research
note (version 9.1) supersede those presented in S06 and V10 and are available
at the MILES website (http://miles.iac.es). This new set of spectra sets the
basis for forthcoming papers describing the extension of our model predictions
to other wavelength ranges.\looseness-2

\begin{acknowledgements}
We would like to thank H. Kuntschner and M. Cappellari for reporting some of the
questions addressed in this research note. We also thank the referee for
very constructive comments. We are grateful to F. Valdes for his assistance
clarifying the spectral resolution of the Indo-US library, and C. Allende Prieto
for useful discussions. JFB and PSB acknowledge support from the Ram\'on y Cajal
Program financed by the Spanish Ministry of Science and Innovation. PSB also
acknowledges an ERC grant within the 6th European Community Framework Programme.
This research has been supported by the Spanish Ministry of Science and
Innovation (MICINN) under grants AYA2009-10368, AYA2010-21322-C03-02 and
AYA2010-21322-C03-03. This paper is based on observations obtained at the Isaac
Newton Telescope, operated by the Isaac Newton Group in the Spanish Observatorio
del Roque de los Muchachos of the Instituto de Astrof\'isica de
Canarias.\looseness-2
\end{acknowledgements}


\listofobjects

\end{document}